# CHI2000 Proposal for Workshop 11:
# The What, Who, Where, When, Why and How of Context-Awareness

By George Tsibidis [♣], Theodoros N. Arvanitis & Chris Baber
Kodak/Royal Academy Educational Technology Research Group,
School of Electronic and Electrical Engineering,
The University of Birmingham
Edgbaston, Birmingham, B15 2TT, England

**ABSTRACT**
The understanding of context and context-awareness is very important for the areas of handheld and ubiquitous computing. Unfortunately, at present, there has not been a satisfactory definition of these two concepts that would lead to a more effective communication in human-computer interaction. As a result, on the one hand, application designers are not able to choose what context to use in their applications and on the other, they cannot determine the type of context-awareness behaviours their applications should exhibit. In this work, we aim to provide answers to some fundamental questions that could enlighten us on the definition of context and its functionality.

**INTRODUCTION**
Over the last few years, many attempts have been made towards defining successfully the *context* or the implicit situational information transferred to a computer by a human during a human-computer interaction [1]. Unfortunately, context-awareness is still not well understood and techniques for developing applications based on context-awareness are in the early stages [1]. In this proposal, we will concentrate on defining the concept of context and we will describe briefly ongoing work that uses elements from context-aware applications [10].

**THE PERCEPTION OF CONTEXT IN HUMAN-COMPUTER INTERACTION**
It is true that a human-to-human communication is much more successful than a human to computer. The reasons lie in the perception of the information passed from one human to another. Humans, usually, possess a common means of realising and understanding situations. By contrast, humans fail to enjoy a similarly effective communication with computers. The reason is that humans do not know what information and how they can transfer it to the computers so that the machines will consider this information as a necessary feedback resulting in a 'good communication'. The question rising is what is the type of information the human should supply the computer with so that it will facilitate the production of more useful services. Before answering this question, it is interesting to look at the features of information a user (human) could send to a computer as part of their interaction. It should be emphasised that in handheld and ubiquitous computing, the information by the user could be changing with the passage of the time leading to a dynamic evolution of the information. This type of behaviour suggests that time should definitely play an essential role in defining the information communicating to the computer. Especially for dynamic systems, the notion of location and the nature and types of activities underlined by the communication are very important and thereby they should be included in a definition regarding the 'information sent'.

Thus, location and identity of the user, activities and operations performed by the system, the time at which a change (for instance, in the location, in the environment, in the types of the activities) constitute some information that can be used to characterise the situation of an entity (person, place, object). This information that is associated to the communication between the user and the computer is called *context* [1]. If an application employs context to provide relevant information and services to the user, it is said to be *context-aware* [1].

The user participating in the human-computer interaction as well as the other humans forming a set, which can be regarded as environment to the user, certainly can benefit from an awareness of context. Especially when the situation of a user or information regarding the system (system is defined as the users, their environment and the applications) changes dynamically, a context-aware application is very important because one can claim that the elements characterising the human-computer interaction do not remain unchanged.

---

[♣] georgiot@eee-fs7.bham.ac.uk

A very plausible question one could raise is whether the context associated to a particular entity plays a more important role and accounts more for a specific human-computer interaction behaviour. A direct answer to this question may be difficult to be obtained because usually, the importance of the information is closely related to what we assume to be important and what goals the computing system aims to achieve. As a result, a general formula as to whose context is important to who or what cannot be derived without taking into account the operation of the system and the relationships between the entities taking part in the system. It is true, there is information that turns out to count more and it should be included in the context.

Certainly, the requirements of a computing system and what we should expect from it determine those essential factors characterising the context of the entities within it. It is known that the information constituting context in a human-computer interaction can be categorised in certain types according to what feature context refers to (*location, identity, activity, time* [1]). The knowledge of the primary context types can help to acquire information residing in other sources of contextual information (*secondary types*). The knowledge of the features of the information underlining context can provide us with the proper tools to develop context-aware applications. In the following section, we are attempting to illustrate a way of implementing context-awareness in applications.

**CONTEXT AS A NETWORK**
In the previous section, we referred to the information that can be context and we proceeded with a classification of the context types. In this section, we are interested primarily in representing this information first as a part of a network where all entities are considered to be nodes and then we are concerned with the details underlining the information. On the other hand, the exchange of information is denoted by links between nodes. The links can be assumed to represent relationships between the entities. The objective this visualisation intends to achieve is to assist in the design of an effective human-computer interaction.

One of the assumptions considered is that there exists a constant dynamic rather than a static relationship between the entities of the network which leads to the emergence of context. The understanding of the context and how it should be used can help in the development of a context-aware application.

**AN EXAMPLE OF CONTEXT-AWARE APPLICATIONS**
After having presenting the theoretical framework where context-aware applications lie, we can proceed with the introduction of an example which could throw some light to some of the above questions set in the previous sections. Actually, it offers a direct answer to the following questions: i. who can benefit from the use of a context-aware application, ii. where a context-aware application can be used, iii. what procedure we should follow in order to build a context-aware application..

This example, also, tends to demonstrate the importance of the existence of context-aware applications.

We report initial development of a personal computer system, which can track visitors to an art gallery or museum and present the visitors with information tailored to their specific requirements, interests and tour [10].

As part of the development process, we need to understand how people can become part of the information network in a museum. This means that visitors move between exhibits, collecting information and making decisions to move onto other exhibits. Their behaviour indirectly affects other museum visitors, e.g., when there are several people wishing to view the same painting, then bunching could occur, leading to an unsatisfactory experience of the museum.

The objective of an effective computing system is to provide the viewer with a 'guided tour' of an art-work. In addition to providing information that relates to the location (the system can know at any time which visitor is in front of what) exhibit, the information can be adapted to match the visitor's level of knowledge. Thus, information can be tailored to the ongoing tour, e.g., with cross-references made to previously encountered exhibits [2]. Previous work has demonstrated the use of adaptive user models in computer-based training [3]. This allowed retrieval of information to be shaped by the users' previous knowledge and experience. Thus, one can define a field of work in which context-aware computing can be used to support and guide visitors. From this perspective, a visit to an art gallery or museum becomes a far more active process (in terms of information and education) than is currently possible. Adapting information for different visitors and supporting interaction with art works can allow people to explore and examine paintings in a deep and complex fashion, which might not be possible without either a formal art education or a well-educated, personal guide.

It should be emphasised that the exhibits of the museum are considered to remain fixed but the people move around, thus lending some dynamic aspect to the network. Understanding the location (and movement) of people within an environment could allow more efficient management of visitors to an exhibition, i.e., minimizing the 'Mona Lisa' effect.

**SMALL-WORLD NETWORKS MODELS**
In order to describe the interaction between visitors and paintings through the relationship underlined by both the prediction of the visitor behaviour and the co-ordination of his/her activities, we will use the concept of Small Worlds. Before we embark on describing in more detail the way this idea is going to be employed, we will explain briefly the concept of Small Worlds.

Any type of network can be represented by a graph, comprising nodes and a set of edges joining the nodes (see Figure 1). The nodes (or vertices) can represent people while the edges may stand for their interpersonal ties. There are two types of networks that have been used extensively towards modelling real-world networks: i. completely ordered (or regular), and ii. completely random. In the former case, every node has the same number of edges and it is connected to its nearest neighbours, forming a highly clustered network. A communication with a node positioned far from the previous one requires the use of a number of vertices lying between them. By contrast, a completely random network is characterised by nodes arbitrarily linked to nodes that can be situated anywhere within the network. As a result, the exchange of information between any two nodes can be achieved with the inclusion of a very small number of intermediate steps. Unlike the ordered network, a completely random network is not clustered.

These networks were originally employed in Physics and Mathematics to represent specific types of problem. Real-world networks appear to lie somewhere between random and regular networks. Such networks are called Small-World Networks (SWN) [4]. SWN are highly clustered while the communication of two arbitrary chosen nodes does not require a large number of intermediate steps. The SWM appears to describe well biological [5], social [6], computer [7], neural [8] and other types of networks.

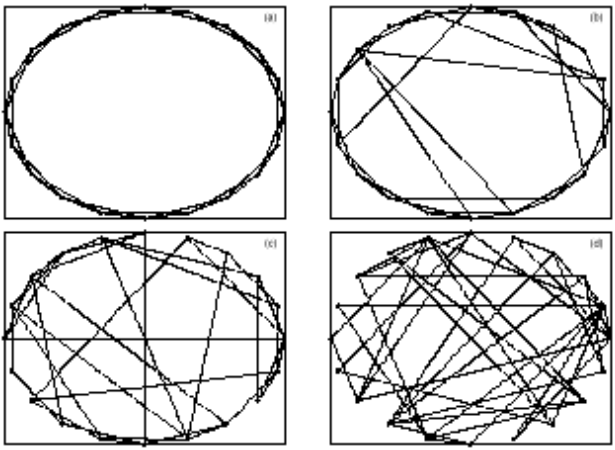

Figure 1: (a). regular network, (d). random network, (b,c) Small World Networks [9]

**THE SMALL-WORLD MODEL AS A DESIGN TOOL OF A CONTEXT-AWARE APPLICATION**

In this study, the employment of the SWN will be used as a design tool that will lead to the optimisation of the network of the people visiting the museum and the network of the information communicating to the visitors. The objective of the network of the information communicating to the visitors is related to achieving the congestion control. At first, the system decides according to the requirements set by the museum how many visitors should stand in front of a certain art-work. Figure 2 shows the results of a short field study investigating 50 visitors to a small gallery. Notice the variation in time that the visitors spend in front of an art-work and in the gallery. This implies variation in visitor activity, which we believe can be captured through time-based measures. For example, if we use 20 minutes as a cut-off, we can divide the sample into approximately two groups in terms of how long they spend looking at an art-work (on average).

If we can determine the average time that a type of visitor will spend looking at a painting, this has implications for how much information to present and how quickly the visitor will move through the gallery. Consequently, our model will need to have time-based measures defining amount of time spent at a painting. Such a measure could also assist in defining 'bunching', e.g., the museum could define a set number of people that can comfortably view a painting (depending on the size, position etc. of the painting), and if the number of people exceeds this limit there will be congestion. In such a situation, the system guides the visitor according to his/her preferences (style of the artifact) to another painting matching the interests of the visitor (thick line in Figure 3) but different from that which would cause congestion. There exists, however, a possibility of more than a 'reasonable' number of visitors guided to the same painting. The local network associated to a certain painting that provides information (thin line in Figure 3) relevant to the painting such as the life of the artist, what inspired the artist, etc. could modify the number of people that part of the museum can accommodate. Thus, in essence, the presence of points (the red squares in Figure 3) that although are related to a painting but reduce the congestion around it relieves the museum.

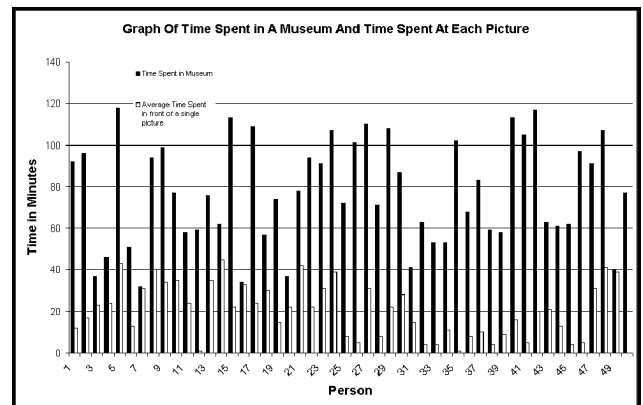

Figure 2: Time-based measures of a visit to a gallery

In Figure 3, both networks are realised in a single network comprising two parts: (i). a local network describing the paintings in the art gallery and (ii). a local network representing the information associated to each painting. The proposed system aims at controlling and reducing the congestion in front of the paintings of the gallery (and thereby avoiding Mona-Lisa type effects). As a result, it would make the visit to a museum more enjoyable.

The question rising is how the application of a SWM could help to build a design tool that would enable the system to direct the visitors of the museum in an effective way minimising the congestion.

The equivalent of the extreme case of a totally random network of paintings would be scenario of people without 'common interests' moving arbitrarily in the museum. The absence of 'common interests' could be easily identified with the non-existence of a system that would take into account what the visitors would like to see and direct them accordingly. In addition, it should be assumed that the visitors do not have certain preferences of what they want to see.

On the other hand, a total random network would be one which is characterised by a large number of people having the same interests and wanting to see at the same time a particular painting. As a result, the time required so that the visitors will enjoy a piece of art increases and so does the congestion.

The SWM is a network that lies in between the above two cases and this is the reason why it is assumed that it could provide a solution to the problem: on the one hand, it accepts that the visitors possess similar properties (they have 'common interests') and the system directs them randomly, however, it takes into account the 'common interests' feature.

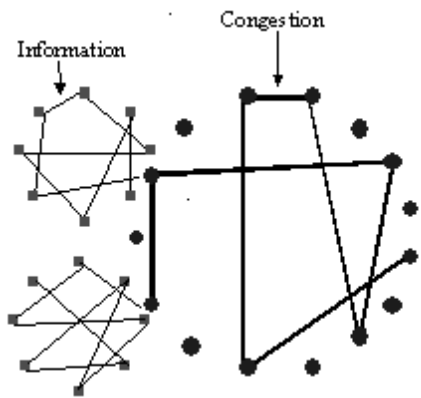

Figure 3: the blue circles ( ● ) represent the paintings of the museum while the red boxes ( ■ ) the information assigned to every painting.

## CONCLUSIONS

In this work, we have described briefly the importance of equipping computer applications with context awareness. The existence of context-aware applications benefits the user since it facilitates the human-computer interaction and it is particularly important for applications supported by handheld and ubiquitous computing where the user's context changes rapidly. We answered some regularly posed questions the knowledge of which could assist us in acquiring a better understanding of what information and how it should be transferred to a computer by a human user.

In an effort to demonstrate the benefits from the use of context-aware applications, we described ongoing work based on a system aiming at tracking and guiding visitors to an art gallery or a museum. The basic elements of the model are developed and the designing of the system is implemented by means of the concept of Small Worlds.

## ACKNOWLEDGEMENTS
G.T. would like to thank Evi Chryssafidou for stimulating discussions.